\documentclass[a4paper]{jpconf}
\usepackage{graphicx}
\usepackage{lineno}
\begin{document}
\title{Liquid Xenon detector R\&D for $0\nu2\beta$ search (KamXP)}

\author{Kota Ueshima}

\address{Research Center for Neutrino Science, Tohoku University, Sendai 980-8578, Japan}

\ead{ueshima@awa.tohoku.ac.jp}

\begin{abstract}
 The R\&D for a new type of liquid xenon (LXe) detector is ongoing to search for the neutrinoless double-beta decay ($0\nu2\beta$). As a result of the KamLAND-Zen experiment, it is very important to realize the all active region detector for the energy deposition of the radiation. Newly developed additional BG reduction techniques will improve the sensitivity of $0\nu2\beta$ search in the future. Our detector concept is LXe stored
 in a new type of plastic scintillator vessel. The wavelength of LXe
 scintillation light is shifted to visible light from 175nm (VUV) on the inner surface of
 the vessel. Therefore, the LXe scintillation light can be detected by photon
 sensors which are far away from the $0\nu2\beta$ target nuclei. The pulse shape 
difference between LXe and plastic scintillator is used for the additional BG
 reduction. In the future, $^{8}$B solar neutrino events will be one of the dominant BGs in
 $0\nu2\beta$ search. To reduce the $^{8}$B solar neutrino BG, the directional information of the Cherenkov light might be useful. The status of LXe detector R\&D to improve the sensitivity of $0\nu2\beta$ search is reported.
\end{abstract}

\section{Introduction}

 By the discovery of the atmospheric neutrino oscillation[1], it was found that the neutrino had finite mass.
However, the neutrino mass scale is still not understood. If $0\nu2\beta$ signal is observed,
there are very important impacts for particle physics, such as the effective neutrino mass, mass hierarchy determination and evidence of Majorana particle.
In addition, the $0\nu2\beta$ is a lepton number violating process.

 There are many $0\nu2\beta$ search experiments in the world using various detectors and BG reduction techniques. We started R\&D to improve the sensitivity for $0\nu2\beta$ search, so called Kamioka double beta decay search using LXe stored in plastic scintillator vessel (KamXP).
The current main BGs of $0\nu2\beta$ search in the KamLAND-Zen experiment[2] are $^{214}$Bi, $^{10}$C and $2\nu2\beta$ tail events. 
Usually $^{214}$Bi events are tagged using Bi-Po continuum decay. However, in case that the $\alpha$-particle emitted by the $^{214}$Po decay stops in the mini-balloon film which is a dead region for the energy deposition by radiation, the $^{214}$Bi events become BG for the $0\nu2\beta$ search. Using a plastic scintillator vessel, there is no dead region for the energy deposition by radiation. Therefore, the $^{214}$Bi events are reduced by the tagging of Bi-Po continuum decay. $^{10}$C is the spallation product caused by cosmic ray muons. In LXe there is almost no carbon, so the BG of $^{10}$C could be reduced drastically. In addition, the diameter of the plastic scintillator vessel is only 90\,cm even if 1\,ton LXe is stored. Therefore, a very compact volume is realized to search for $0\nu2\beta$.

 The pulse shape discrimination (PSD) study of LXe was performed for direct dark matter searches [3-6].
In the present R\&D, the PSD between LXe and plastic scintillator is used for the additional BG reduction to improve the sensitivity of $0\nu2\beta$ search.

\section{Plastic scintillator vessel development}

 The wavelength of LXe scintillation light is 175nm (VUV). To detect a light signal with PMTs which are far away from the LXe target, the wavelength of the LXe scintillation light has to be shifted to visible light[7]. Tetra-phenyl butadiene (TPB) was used as a wavelength shifter in the present work. We developed new types of TPB doped plastic scintillator vessel using various plastics. The base of normal plastic scintillator is polystyrene. At first, 4\,cm diameter small vessel was developed using 3\,wt\% TPB doped in normal plastic scintillator. Figure\,1 shows the pulse shape difference between wavelength shifted LXe scintillation light and plastic scintillator scintillation light. The organic solvent resistance of the polystyrene is very weak. Therefore the polystyrene based plastic scintillator can not be used in organic liquid scintillator. 
 
  In the present R\&D two types of plastic were used for vessel development. Polyethylene naphthalate (PEN) and Tritan$^{TM}$ have excellent properties for organic solvent resistance, mechanical strength and low temperature. Figure\,2 shows the developed new plastic scintillator vessels. Figure\,3 shows the Tritan$^{TM}$ based of the vessel doped with 2\,wt\% of TPB. The diameter of the vessel is 20\,cm. About 10\,kg LXe could be stored in the vessel.

% Figure
\begin{figure}[h]
\begin{center}
\includegraphics[scale = 1.0]{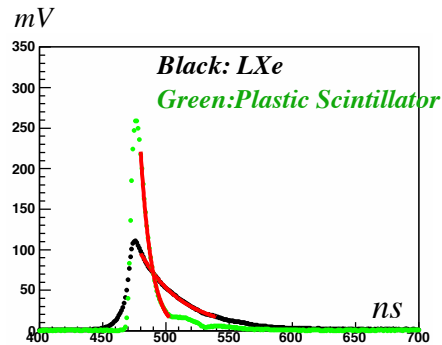}
\caption{The average scintillation light waveform of LXe and plastic scintillator in the selected same charge range.}
\label{fig:7}
\end{center}
\end{figure}

% Figure
\begin{figure}[h]
\begin{minipage}{18pc}
\begin{center}
\includegraphics[width=14pc]{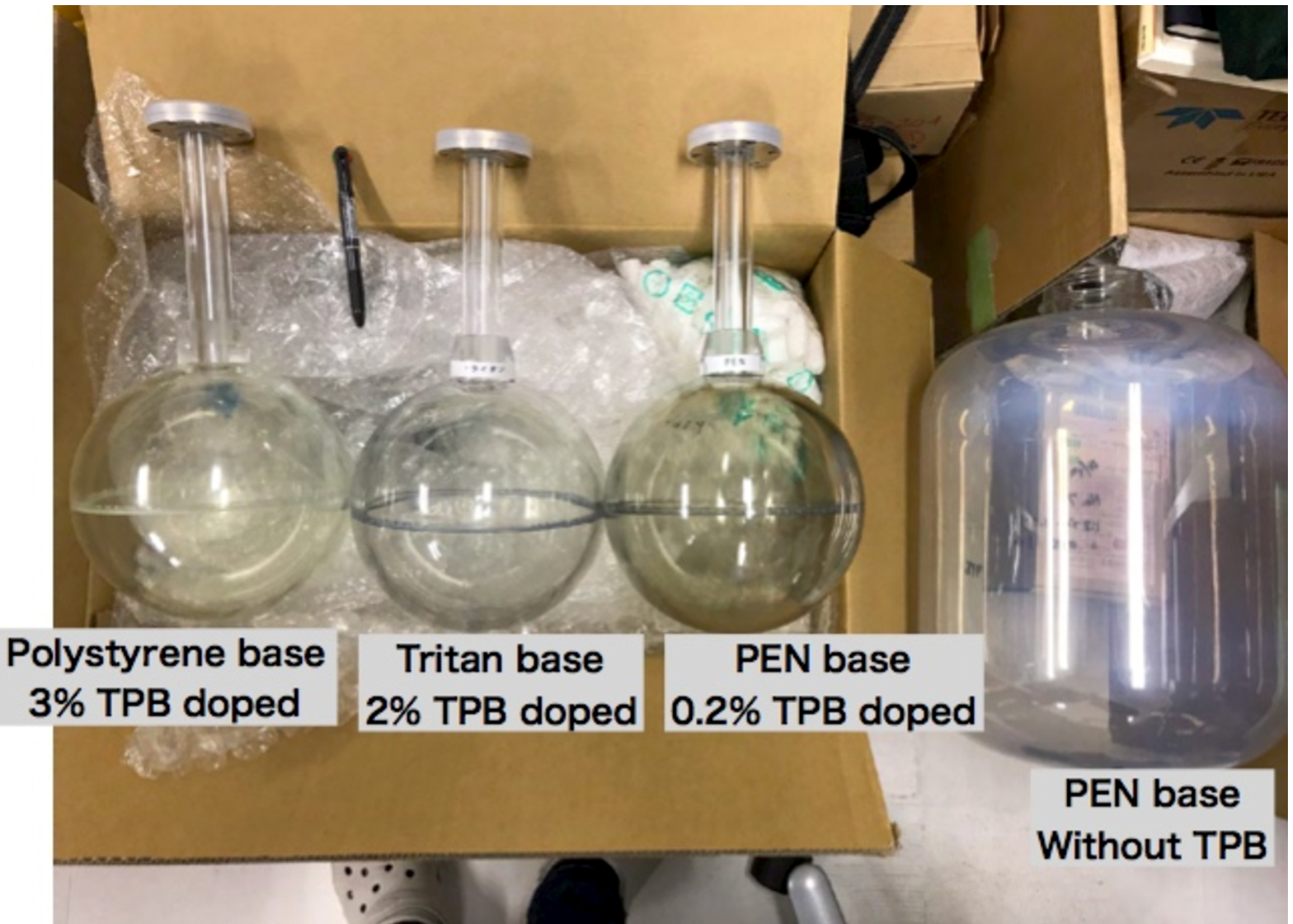}
\caption{The developed new plastic scintillator vessels.}
\end{center}
\end{minipage}\hspace{2pc}%
\begin{minipage}{18pc}
\begin{center}
\includegraphics[width=14pc]{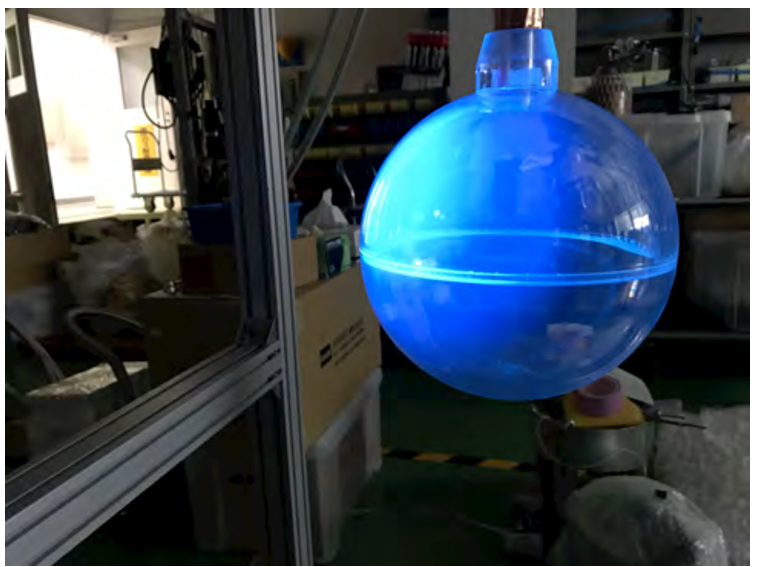}
\caption{\label{ves} Tritan$^{TM}$ based plastic scintillator vessel.}
\end{center}
\end{minipage} 
\end{figure}

\section{Cherenkov ring detection method and R\&D setup}

  The extraction of the Cherenkov light component from the detected total light waveform is very difficult. The Cherenkov light component contains only a few\,\% of the organic liquid scintillator scintillation light[8].
  
  In this work the LXe scintillation light is absorbed on the inner surface of the plastic scintillator vessel and re-emitted. In contrast, the Cherenkov light is
 passing through the plastic scintillator vessel and detected by PMTs directly.
 Therefore, the rise time of the waveform of LXe scintillation light might be slow as shown in Figure\,1.
 Using the rising component of the waveform, the Cherenkov light component in LXe
 scintillation light might be detected effectively. The directional information will be extracted from LXe
 scintillation light using the Cherenkov ring information. 
 
 Figure\,4 shows the detector setup to detect Cherenkov light. The detector consists of the plastic scintillator vessel, transparent adiabatic vacuum vessel and PMTs. Seven 2-inch squared 2x2 multi anode PMTs (R12699) and two 2-inch PMTs (R6041) were used for Cherenkov light detection as shown in Figure\,5. The typical transit time spread value (FWHM) is 0.41\,ns (R12699) and 0.75\,ns (R6041), respectively. 
In addition, to increase the light yield of the wavelength shifted LXe scintillation light, 8-inch PMT (R5912) and 2-inch PMT were also mounted at the bottom and the equator. The total 32\,channels of PMT signals were measured by 2 FADCs (CAEN DT5742B). The sampling frequency of the waveform could be chosen from 750\,MS/s to 5\,GS/s.
At this time about 1\,kg LXe was installed in the plastic scintillator vessel to test the LXe stability in the cryostat.
The LXe temperature and pressure could be controlled at 170\,K and 0.035\,MPaG, respectively. 
 
\begin{figure}[h]
\begin{minipage}{18pc}
\begin{center}
\includegraphics[width=14pc]{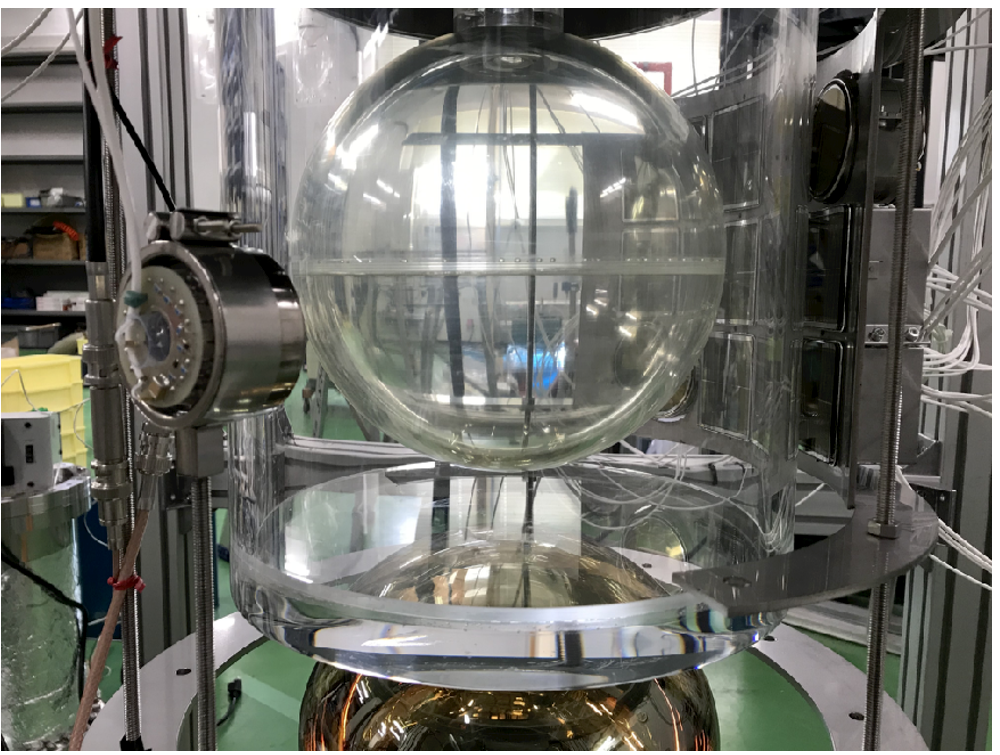}
\caption{\label{setup} Cryostat for Cherenkov light detection. The $\phi$ 20\,cm plastic scintillator vessel and transparent adiabatic vacuum vessel. }
\end{center}
\end{minipage}\hspace{2pc}%
\begin{minipage}{18pc}
\begin{center}
\includegraphics[width=12pc]{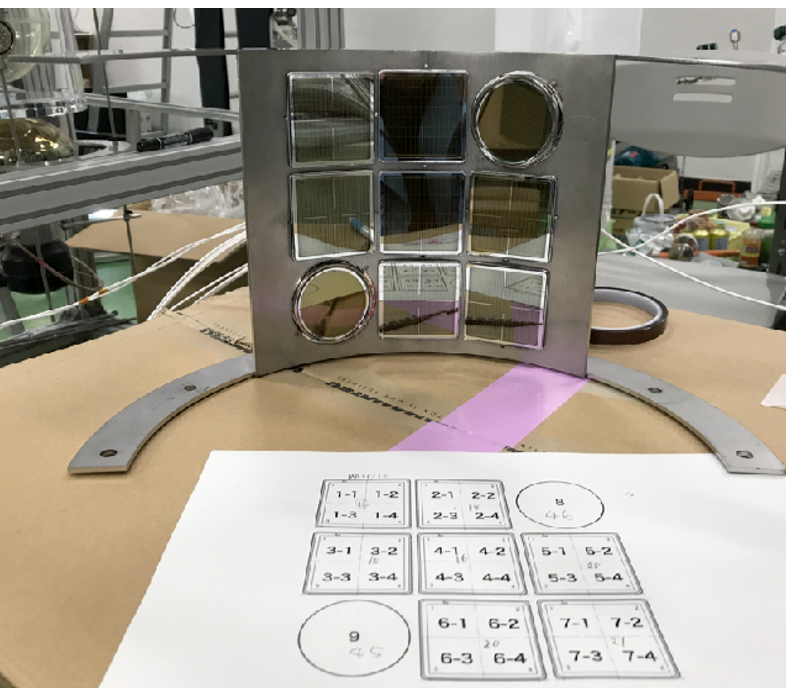}
\caption{\label{pmt} PMT array for Cherenkov light detection. }
\end{center}
\end{minipage} 
\end{figure}

\section{Current status}

 The LXe scintillation light was measured using a $^{60}$Co source. The wavelength shifted scintillation light was observed using 1\,kg LXe. In addition, the preliminary PSD performance between plastic scintillator and wavelength shifted LXe scintillation was evaluated using 8-inch PMT data. The {\it PSD ratio} was defined by the prompt scintillation light detected in the first 20\,ns divided by the total amount of the scintillation light. Figure\,6 shows the preliminary PSD performance of before and after the 1\,kg LXe filled into the Tritan$^{TM}$ based vessel.
The amount of LXe will be increased to about 10\,kg this year. The extraction efficiency of the directional information from the detected light waveform will be evaluated.   

% Figure
\begin{figure}[h]
\begin{center}
\includegraphics[scale = 0.7]{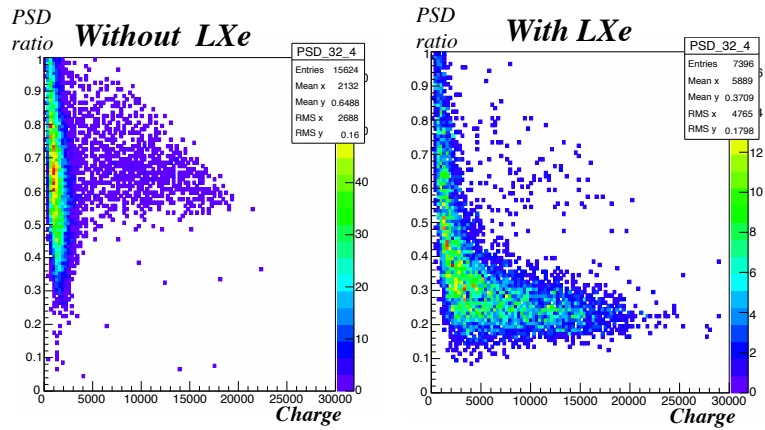}
\caption{Preliminary PSD performance between wavelength shifted LXe scintillation and Tritan$^{TM}$ based plastic scintillator.}
\label{fig:1}
\end{center}
\end{figure}

\section{Summary}

  The status of LXe detector R\&D to improve the sensitivity of the $0\nu2\beta$ search is reported.
To realize the all active region detector for the energy deposition of radiation, the new type of plastic scintillator vessels were developed.
In addition, to reduce $^{8}$B solar neutrino BG using directional information of Cherenkov light, the Cherenkov light detection setup was developed.  
The PSD performance between LXe and plastic scintillator was studied using 1\,kg LXe.
We will evaluate the extraction efficiency of the directional information using the Cherenkov light component at the 2.5\,MeV signal region. 

\section*{Acknowledgments}
This work is supported by the Japan Society for the Promotion of Science (JPSP) Grant-in-Aid for Young Scientists (A), Project Number 17H04834.

\section*{References}

\medskip

\numrefs{99}
\item Y.Fukuda {\it et al}. {\it Phys. Rev. Lett.}  \textbf{81} (1998) 1562
\item A.Gando {\it et al}. {\it Phys. Rev. Lett.}  \textbf{117} (2016) 082503
\item R.Bernabei {\it et al}.  {\it Phys. Lett. B} \textbf{436} (1998) 379
\item G.J.Alner {\it et al}. {\it Astropart. Phys.} \textbf{23} (2005) 444
\item J.Kwong {\it et al}. {\it Nucl. Instr. and Meth. A} \textbf{612} (2010) 328
\item K.Ueshima {\it et al}. {\it Nucl. Instr. and Meth. A}  \textbf{659} (2011) 161
\item K.Ueshima {\it et al}. {\it Nucl. Instr. and Meth. A}  \textbf{594} (2008) 148
\item C.Aberle {\it et al}. {\it JINST}  \textbf{9} (2014) P06012

\endnumrefs

\end{document}